\documentclass[11pt]{article}
\pdfoutput=1

\usepackage{amsmath}
\usepackage{amsfonts}
\usepackage{amssymb}
\usepackage{graphicx,color}
\usepackage{subfigure}
\usepackage{hyperref}
\usepackage[usenames,dvipsnames,svgnames,table]{xcolor}
\usepackage{enumerate}

\usepackage[left=2cm,top=2.5cm,right=2cm,bottom=2cm]{geometry}
\def \be  {\begin{equation}}
\def \ee  {\end{equation}}
\def \ba  {\begin{eqnarray}}
\def \ea  {\end{eqnarray}}

\def \NI {{\it NI}}
\def \HNI {{\it HNI}}
\def \WNI {{\it WNI}}

\begin{document}

\title{Constraining Hybrid Natural Inflation with recent CMB data} 
\author{J. Alberto V\'azquez$^{a}\,\!\!$, Mariana Carrillo--Gonz\'alez$^{b,c}\,\!$, Gabriel Germ\'an$^{d}$,\\
Alfredo Herrera--Aguilar$^{e,f,g}\,\!\!$, Juan Carlos Hidalgo$^{d}$\\
{\normalsize \textit{$^a$Brookhaven National Laboratory, 2 Center Road,
Upton, NY 11973, USA.}}\\
{\normalsize \textit{$^b$Perimeter Institute for Theoretical Physics,}}\\
{\normalsize \textit{31 Caroline St. North, Waterloo ON N2L 2Y5, Canada.}}\\
{\normalsize \textit{$^c$Department of Physics and Astronomy,}} 
{\normalsize \textit{University of Waterloo,}}\\
{\normalsize \textit{200 University Av. West, Waterloo, ON N2L 3G1, Canada}}\\
{\normalsize \textit{$^d$Instituto de Ciencias F\'isicas,} }
{\normalsize \textit{Universidad Nacional Aut\'onoma de M\'exico,}}\\
{\normalsize \textit{Apdo. Postal 48-3, 62251 Cuernavaca, Morelos, M\'{e}xico.}}\\
{\normalsize \textit{$^e$Departamento de F\'{\i}sica, Universidad Aut\'onoma Metropolitana Iztapalapa,} }\\
{\normalsize \textit{San Rafael Atlixco 186, CP 09340, M\'exico D. F., M\'exico.}}\\
{\normalsize \textit{$^f$Instituto de F\'{\i}sica y Matem\'{a}ticas,}}
{\normalsize \textit{Universidad Michoacana de San Nicol\'as de Hidalgo,}}\\
{\normalsize \textit{Edificio C--3, Ciudad Universitaria, CP 58040, Morelia, Michoac\'{a}n, M\'{e}xico.}}\\
{\normalsize \textit{$^g$Instituto de F\'{\i}sica,
	 Benem\'erita Universidad Aut\'onoma de Puebla,}}\\
{\normalsize \textit{ Apdo. postal J-48, 72570, Puebla, Puebla, M\'exico.}}}
\date{}
\maketitle

\begin{abstract}
We study the {\it Hybrid Natural Inflation} ({\it HNI}) model and some of its realisations in the light of recent CMB observations, 
mainly Planck temperature and WMAP-9 polarization, and compare with the recent release of BICEP2 dataset.  The inflationary sector of {\it HNI} is essentially given by the potential $V(\phi) = V_0\left(1+a\cos \left(\frac{\phi}{f} \right) \right)$, where $a$ is a positive constant smaller or equal to one and $f$ is the scale of (pseudo Nambu-Goldstone) symmetry breaking. We show that to describe the {\it HNI} model realisations we only need two observables; the spectral index $n_s$, the tensor-to-scalar ratio, and a free parameter in the amplitude of the cosine function $a$. We find that in order to make the {\it HNI} model compatible with the BICEP2 observations, we require a large positive running of the spectra. We find that this could over-produce primordial black holes (PBHs) in the most theoretically consistent case of the model. This situation could be alleviated if, as recently argued, the BICEP2  data do not correspond to primordial gravitational waves.

\end{abstract}

\section{Introduction} \label{Intro} 

\noindent 

Cosmological inflation \cite{Guth:1981,Linde:1982,Albrecht:1982,Lyth:1998xn} not only explains the homogeneity of the universe on very large scales, but provides a theory for the seeds 
of structure, explaining the observed level of anisotropy. Typical 
slow-roll models generically predict Gaussian, adiabatic and nearly scale-invariant primordial fluctuations, but it is 
ultimately observations of the galaxy distribution and the microwave background anisotropies which determine the shape of the 
primordial spectra. The simplest assumption is that the initial spectrum has a power-law form, parameterised in terms of the 
spectral amplitude $A_i$ and the spectral indices $n_i$, where the {\it{i}} refers to scalar or tensor components. Recent analyses 
have shown, however, that the spectral index may deviate from a constant value (close to unity) and so consideration of models 
that provide some running of the index is warranted \cite{Vazquez11, Vazquez12}.

The inflationary sector of the models explored in this paper is given by the potential
\begin{equation}
V(\phi) = V_0\left(1+a\cos \left(\frac{\phi}{f} \right) \right),
\label{PotHNI}
\end{equation}
where $a$ is a positive constant less or equal to one and $f$ is the scale of (pseudo Nambu-Goldstone) symmetry breaking. 
 \NI~ was originally proposed 
to justify the inflationary potential in the context of high energy physics, where the inflaton is a pseudo Nambu-Goldstone boson 
resulting from symmetry breaking \cite{Freese:1990rb,Freese:1993,Freese:2008if}. However, 
as we also show in this paper, when tested against observations, $f$ lies above the Planck scale. This is usually attributed to 
the fact that the same inflaton must end inflation after at least 50 e-folds of expansion. 

To overcome this possible weakness in the \NI~ model, other, more complete and equally well motivated models have been 
formulated. In particular, the \textit{Hybrid Natural Inflation} (\HNI) model was first presented in \cite{Ross:2009hg,Ross:2010fg}. The model is an improvement over \NI~in that a waterfall-like auxiliary field \cite{Linde:1994} is 
considered to end inflation thus allowing the scale $f$  to get Planckian or sub-Planckian values. It is thus imperative to constrain the parameter space of the model with the observations at hand. 
In this paper we derive 
constraints for the parameters  of the \HNI~ model when subject to the latest CMB observations. We 
do this for two cases: one with no priors on the model parameters, and a second case where the  symmetry breaking scale $f$ 
takes the Planck value. 

In an effort to restore realistic values of the symmetry breaking parameter $f$, we relieve the inflaton in \NI~  from the 
requirement of ending inflation, and endow the model with a second, waterfall scalar field,  responsible of the end of 
inflation (see \cite{Ross:2009hg} for the details). In fact, if we set the parameter $a = 1$ in Eq.~\eqref{PotHNI}, we recover a hybrid version of \textit{Natural Inflation}(\NI),  we call this scenario \textit{Waterfall Natural Inflation}(\WNI). With the constraints imposed by observations at hand, we explore the theoretical and phenomenological  differences between \WNI~ and \HNI.  Throughout this paper we carry two parallel analyses. A first analysis only taking the Planck data on account, with the WMAP polarisation data \cite{WMAP9}, and a second analysis taking the BICEP2 polarisation data as a product of primordial gravitational waves, as reported in the collaboration paper \cite{Ade:2014xna}. Alternative interpretations of these results, attributing the polarisation signal to foreground dust are still to be confirmed by the Planck satellite and hence the motivation to contrast both datasets studied here.

Our presentation is organized as follows: in Section~\ref{Frame} we present both models of inflation, {\it HNI} and \WNI, identifying the inflationary stage and the
form of their spectral parameters. In Section~\ref{Results} we present  likelihood contours of parameter values, in light of the latest Planck data release \cite{Ade:2013uln} and WMAP  polarization maps (WP) \cite{WMAP9}, hereafter Planck+WP; and separately for the combination of Planck+WP and the results of the BICEP2 experiment \cite{Ade:2014xna} (here Pl+ WP+BICEP2. We do this for three different cases: The general {\it HNI} model with no priors on the pair of parameters $\{a,f\}$, the case $a=1$ corresponding to \WNI, and finally  the case with no prior over $a$ but with the symmetry breaking scale $f$ restricted to the largest physical value $f = 1$, in Planck units. In Section~\ref{Analyses} we analyse the featured results identifying pathological cases where either the scale $f$ acquires unphysically large values, or where the running could over-produce primordial black holes, as is the case for other models of inflation where the potential slope flattens toward the end of inflation \cite{Kohri:2007qm,Alabidi:2009bk}. 
Finally, in Section~\ref{Conclu} we conclude by discussing the observational and theoretical constraints to the featured models.

\section{Slow-roll parameters and observables} \label{Frame} 

\subsection{Slow-roll parameterisation of spectra}\label{Resultsb} 

\noindent

In slow-roll inflation, the spectral indices are given in terms of the slow-roll parameters of the models, which involve the potential $V(\phi)$ and its derivatives (see e.g. \cite{Liddle:94},
\cite{Liddle:2000cg})
\begin{equation}
\epsilon \equiv \frac{M^{2}}{2}\left( \frac{V^{\prime }}{V }\right) ^{2},\quad
\eta \equiv M^{2}\frac{V^{\prime \prime }}{V}, \quad
\xi_2 \equiv M^{4}\frac{V^{\prime }V^{\prime \prime \prime }}{V^{2}},\quad
\xi_3 \equiv M^{6}\frac{V^{\prime 2 }V^{\prime \prime \prime \prime }}{V^{3}}.
\label{Slowparameters}
\end{equation}%
Here primes denote derivatives with respect to the inflaton $\phi$ and $M$ is the
reduced Planck mass $M=2.44\times 10^{18} \,\mathrm{GeV}$ which we set
$M=1$ in most of what follows.
In the slow-roll approximation observables are given by (see e.g.  \cite{Ade:2013uln},\cite{Liddle:2000cg})
\begin{eqnarray}
n_{t} &=&-2\epsilon =-\frac{r}{8} , \label{Int} \\
n_{s} &=&1+2\eta -6\epsilon ,  \label{Ins} \\
n_{tk} &=&\frac{d n_{t}}{d \ln k}=4\epsilon\left( \eta -2\epsilon\right), \label{Intk} \\
n_{sk} &=&\frac{d n_{s}}{d \ln k}=16\epsilon \eta -24\epsilon ^{2}-2\xi_2, \label{Insk} \\
n_{skk} &=&\frac{d^{2} n_{s}}{d \ln k^{2}}=-192\epsilon ^{3}+192\epsilon ^{2}\eta-
32\epsilon \eta^{2} -24\epsilon\xi_2 +2\eta\xi_2 +2\xi_3, \label{Inskk} \\
A_s(k) &=&\frac{1}{24\pi ^{2}}\frac{\Lambda^4}{%
\epsilon _H},
\label{IA} 
\end{eqnarray}
where $n_{tk}$ denotes  the running of the tensor index, $n_{sk}$ the running of the scalar index and $n_{skk}$ the running of the running, in a self-explanatory notation. The density perturbation at wave number $k$ is $A_s(k)$, the scale of inflation is $\Lambda$ with $\Lambda \equiv V_{H}^{1/4}$ and $r\equiv A_t/A_s$ the ratio of tensor to scalar perturbations. All the quantities with a subindex ${}_H$ are evaluated at the scale $\phi_{H}$, at which the perturbations 
are produced, some $50-60$ e-folds before the end of inflation. In particular the pivot scale is set to $k_H = 0.002\,\mathrm{Mpc}^{-1}$  \cite{Ade:2013uln}.
The limited number of slow-roll parameters in Eq.~\eqref{Slowparameters} yields consistency relations between inflation observables, e.g., 
\begin{eqnarray}
n_{t} &=&-\frac{r}{8} , \label{Slowr} \\
n_{tk} &=& \frac{r}{64}\left(r-8\delta_{ns}\right),  
\label{Slowntk}
\end{eqnarray}
where we simplify by defining $\delta_{ns}\equiv 1-n_s$.

It is a well established convention to assume a parameterised form for the spectra of scalar and tensor perturbations, and the simplest shape is that of a power-law, parameterised in
terms of the spectral amplitude $A_i$ and the spectral indices $n_i$, where the subindex ${}_i$ refers to either scalar $(s)$ or tensor $(t)$ components. Slow-roll inflation 
predicts the spectrum of curvature perturbations to be close to scale-invariant. Based on this result, the primordial power spectra $\mathcal{P}_i$, scalar and tensor,  are commonly expressed as:
\begin{eqnarray}
\mathcal{P}_s(k)&=&A_s \left( \frac{k}{k_H}\right)^{(n_s-1) },\label{ss1}\\
\mathcal{P}_t(k)&=&A_t \left( \frac{k}{k_H}\right)^{n_t}=rA_s\left( \frac{k}{k_H}\right)^{n_t}.
\label{ts1}
\end{eqnarray}
The vast amount of data at hand allows cosmologists to consider a more accurate description of the power spectra \cite{Vazquez13}. This improvement is parameterised by including the running, and even the running of the running of scalar perturbations \cite{Powell12}. In this case, the expressions above become, following \cite{Ade:2013uln},
\begin{eqnarray}
\mathcal{P}_s(k)&=&A_s \left( \frac{k}{k_H}\right)^{(n_s-1) + \frac{1}{2}n_{sk}  \ln\left(\frac{k}{k_H}\right) + 
	       \frac{1}{6} n_{skk} \left(\ln\left(\frac{k}{k_H}\right)\right)^2  },\label{ss2}\\
\mathcal{P}_t(k)&=&A_t \left( \frac{k}{k_H}\right)^{n_t+ \frac{1}{2} n_{tk} \ln\left(\frac{k}{k_H}\right) }\label{ts2},
\end{eqnarray}
all indices are such that, for example, $n_{s}$ stands $n_{s}(k_H)$ implicitly throughout the paper.
Thus we may consider scale dependent indices as,
\begin{eqnarray}
n_{s}(k) -1&\equiv& \frac{d ln \mathcal{P}_s}{d \ln k}=n_{s} -1+n_{sk}\ln\left(\frac{k}{k_H}\right)+\frac{1}{2} n_{skk}\left(\ln\left(\frac{k}{k_H}\right)\right)^2,  \label{Insk2}\\
n_{t}(k) &\equiv& \frac{d ln \mathcal{P}_t}{d \ln k}=n_{t} + n_{tk}\ln\left(\frac{k}{k_H}\right) . 
\label{Intk2}
\end{eqnarray}

Although the power-law spectra of Eqs.~(\ref{ss1}) and (\ref{ts1}) have provided reasonable agreement with cosmological observations, recent analyses of Planck data have shown that if a running of the scalar spectral-index be a free parameter, there exists a preference  for a negative running-value at 2$\sigma$ \cite{Ade:2013uln}. On the other hand, the additional parameters ($n_{sk}, n_{skk}$) do not only help to describe the underlying model more accurately on observable scales, but they also allow for additional shapes of the potential at stages of inflation away from the observable range. In general, we want to avoid positive values of the running as they may result in an over-production of Primordial Black Holes (PBHs) at the smallest scales, near the end of inflation \cite{Kohri:2007qm,Alabidi:2009bk}. We will address this point for the case of \HNI~in Sec.~\ref{Analyses}.

\subsection{Framework for Hybrid Natural Inflation} 
\begin{figure}[t!]\centering
\includegraphics[width=9cm]{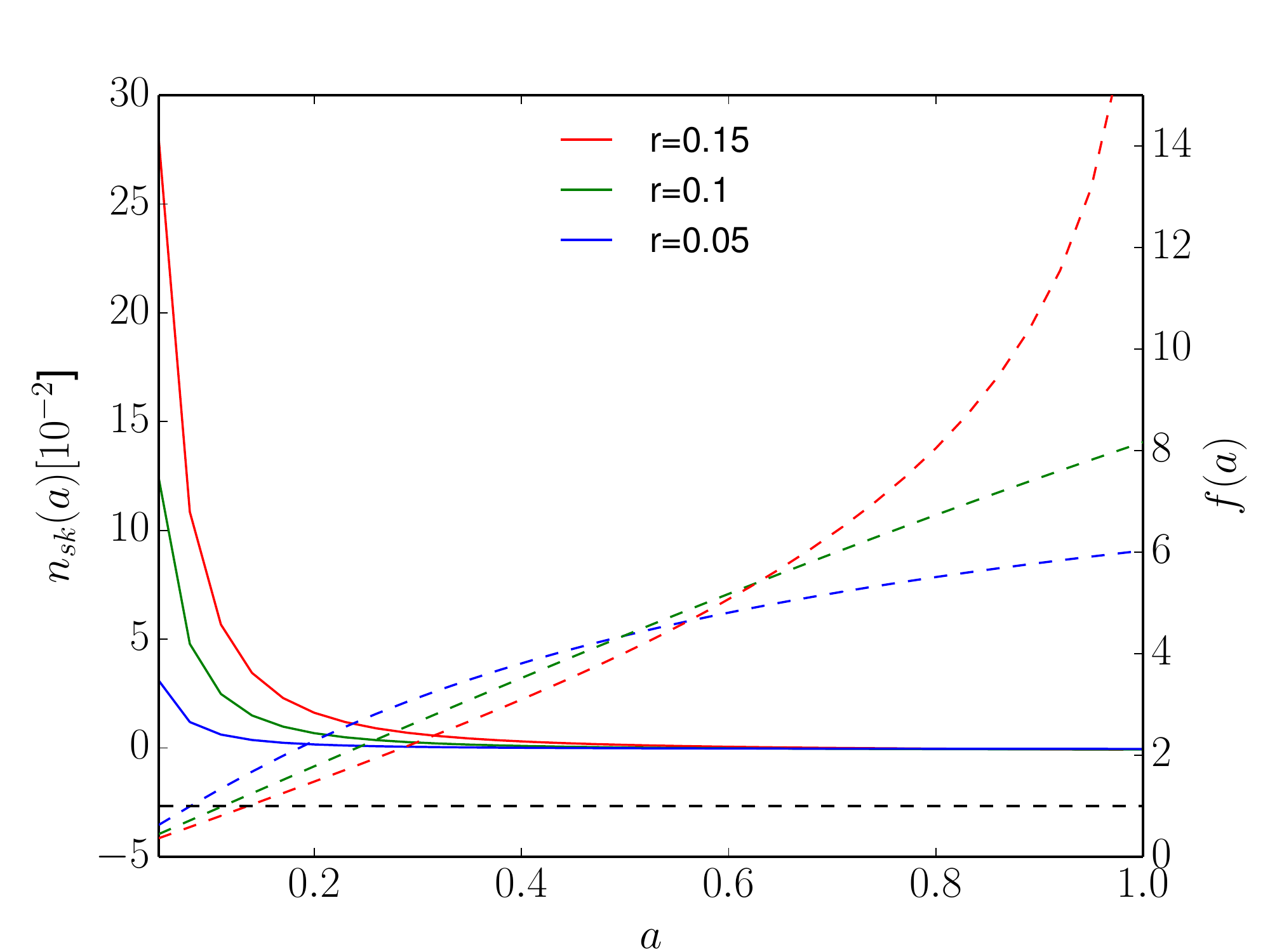}
\caption{\footnotesize{{Plot of $n_{sk}$ (solid lines) and $f$ (dashed lines) as functions of $a$, at fixed $n_s=0.96$ for $r=0.05, 0.1, 0.15$ values,} and as dictated by Eqs.~\eqref{HNInsk} and \eqref{HNIf} respectively. 
Notice that for small $a$ the running becomes large; on the other hand when $a$ approaches 
unity (black dashed line) the running tends to vanish. In this case results are very similar to just considering the case $n_{sk}=0$. We analyse specific cases in Sec. \ref{Results}}.}
\label{PS}
\end{figure}

\noindent

Let us now focus on the inflationary sector of the model given by  the potential of Eq.~\eqref{PotHNI}. The running and running of the running of the scalar power spectrum satisfy the following relations 
\cite{Carrillo-Gonzalez:2014tia}
\begin{eqnarray}
n_{sk}&=&\frac{r}{32}\left(3r-16\delta_{ns}+\frac{8}{f^2} \right),
\label{HNInsk}\\
n_{skk}&=&\frac{r}{128}\left( 3r^2+\frac{12}{f^2}r-32\left( 2\delta_{ns}-\frac{1}{f^2} \right)\delta_{ns}  \right).
\label{HNInskk}
\end{eqnarray}
From Eq.~\eqref{Ins} for the spectral index we obtain an expression for $f$ involving the second parameter of the model, $a$:
\begin{equation}
 f=\frac{4a}{  \left( r\left(1-3 a^{2}\right)+8a^{2}\delta _{ns} +\sqrt{4 a^{2}\left(4\delta _{ns}-r\right)^{2}+r^{2}\left(1-a^{2}\right)} \right)^{1/2}  }.
\label{HNIf}
\end{equation}
It is thus evident that in a slow-roll stage, this model is fully described in terms of the parameter $ a$ alone.
The {\it HNI} model allows for more flexibility in the potential, since the constraint coming from the number of e-folds does not fix $f$. 
The end of inflation is specified by a waterfall field which plays no role during inflation. 
When considering the particular case $a=1$ we are reduced to a \emph{waterfall} version of Natural Inflation.

In Fig.~\ref{PS}, we plot $f$ and $n_{sk}$ in terms of $a$, at a fixed $n_s=0.96$ and for $r=0.05, 0.1, 0.15$ values.
If we assume the values of BICEP2 of the order $r\sim 0.15$ (red lines), for  $a=1$ the running tends to be very close to zero and $f >> 1$. 
On the other hand, if we consider values allowed by Planck (blue lines), $f$ values decreases to order $f\sim 6$.
Similarly for $f=1$ values (black dashed line),  $a$  tends to be order of $a\sim 0.1$ and hence $n_{sk}\sim 0.1 $ for BICEP2
and $n_{sk}\sim 0.01 $ for Planck values (see however subsection \ref{Resultsf},  where we fix $f$ to the value $f=1$ and study its consequences).

From Fig.~1 we can identify the corner of parameter space where a theoretically consistent value for $f$ ($\leq 1$) is obtained during slow-roll. We go further to constrain the \HNI~ model by employing the most recent tracers of the CMB.
In particular, we use the latest release of Planck dataset \cite{Ade:2013uln} which, combined with the nine-year WMAP
 polarisation maps \cite{WMAP9},  allow to constrain the temperature and polarisation spectra (T, E and B components). Furthermore, we provide a joint-analysis with the BICEP2 data release  \cite{Ade:2014xna}, interpreting these results as a non-vanishing tensor contributions. To carry out the exploration of 
the parameter space, we incorporate the \HNI~ potential into the standard cosmological equations, by performing the minor modifications to the {\sc CAMB} code \cite{Lewis:1999bs}. 
We use the CAMB runs to explore the parameter space with the software {\sc CosmoMC} \cite{Cosmo}. 

Throughout the analysis we consider purely Gaussian adiabatic scalar and tensor
perturbations.  We assume a standard
$\Lambda$CDM model specified by the following parameters: the physical
baryon density $\Omega_{\rm b} h^2$ and CDM density $\Omega_{\rm DM}
h^2$, where $h$ is the dimensionless Hubble parameter such that
$H_0=100h$ kms$^{-1}$Mpc$^{-1}$; 
$\theta$, which is $100\times$ the ratio of the sound
horizon to angular diameter distance at last scattering surface, and the
optical depth $\tau$ at reionisation.

\section{Results: Cosmological constraints} \label{Results} 

\noindent

In this section we present the one- and two-$\sigma$ likelihood contours for the set of spectral parameters $\{n_s,a,f,n_{sk},n_{skk}\}$\footnote{We omit plotting the tensor index $n_t$ because in the assumed slow-roll regime it is linearly proportional to the tensor-to-scalar ratio $r$, cf. Eq.~\eqref{Int}.}, plotted against the tensor to scalar ratio $r$, in three different cases of the model in Eq.~\eqref{PotHNI}. In Sec.~\ref{Resultsaf} we study the case when the parameters $\{a,f\}$ are allowed to take any constant value in the range $0 < a < 1$ and $f > 0$; In Sec.~\ref{Resultsa} we look at the case of \WNI~, where we fix $a= 1$ \textit{a priori}; and In Sec.~\ref{Resultsf} the case of $f = 1$ is studied. This is justified as a \emph{just} realistic value for the symmetry breaking parameter $f$. 

\subsection{Full Hybrid Natural Inflation} \label{Resultsaf}

\noindent

\begin{figure}[t!]\centering
\includegraphics[width=17cm,height=5.5cm]{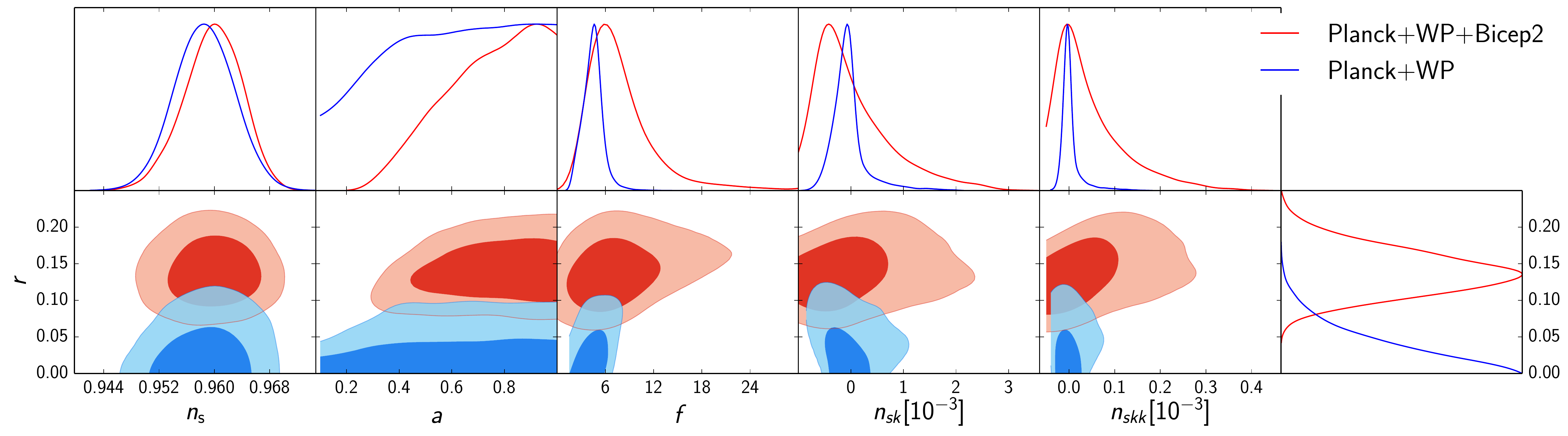}
\caption{\footnotesize{We show results for the full hybrid natural inflation model where $f$ is related to $a$ through Eq.~(\ref{HNIf}). 
We show 1D and 2D posterior marginalized distributions of two sampling parameters 
$\{n_s, r\}$ and derived parameters $\{a,f, n_{sk}, n_{skk}\}$ using Planck+WMAP dataset (in blue), and the combination
of both datasets (PL+WP+BP2, in red).  }}
\label{figHNIaf}
\end{figure}

As introduced above, the \HNI~model features $a$ as a free parameter taking any value in the interval $0<a <1$. With the scale $f$ allowed to take any given positive value, Fig.~\ref{figHNIaf} shows the likelihood contours which present no degeneracy between parameters. From the figure we can read a small running consistent with $n_{sk}=0$, both in the analysis using Planck+WP data only 
and the combined Planck+WP+BICEP2 data. The results suggest that, when leaving $f$ arbitrary, {\it HNI} shows no significant improvement over {\it WNI} since $f$ has to be bigger that 
one and the running uninterestingly small. More interestingly, the Planck+WP+BICEP2 data favour a vanishing running but with a parameter $a$ close to one ($0.8$) and $f \approx 8$ as shown 
in Table~\ref{table1}, instead of the small and always negative running of {\it WNI} with $a=1$ and larger $f$  \cite{Freese:2014nla}. The preferred value of $a$ approaching unity is close to the standard \NI~ model.

\subsection{Hybrid Natural Inflation with $a=1$} \label{Resultsa} 

\noindent

\begin{figure}[t!]\centering
\includegraphics[width=18cm,height=6.0cm]{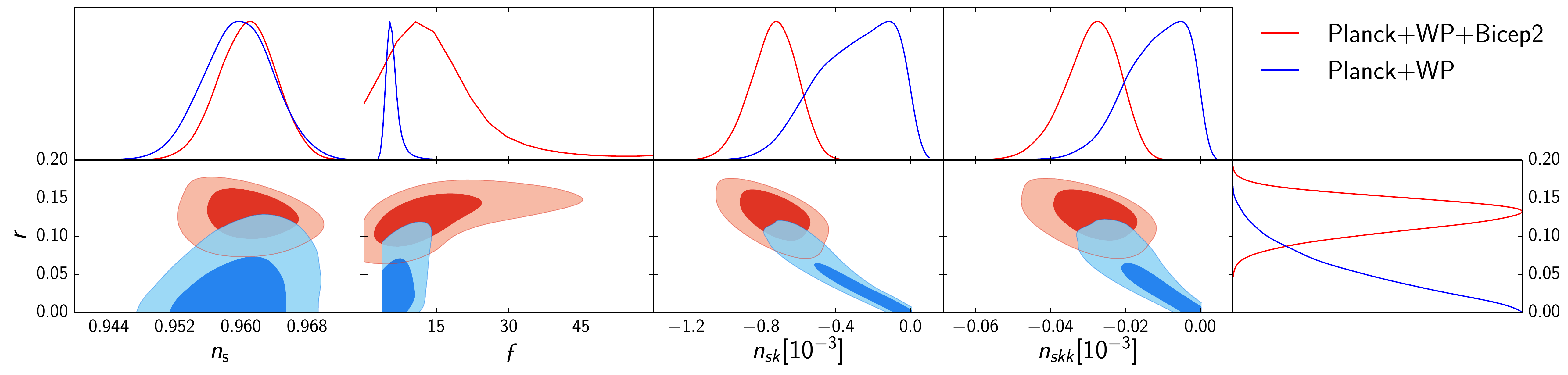}
\caption{\footnotesize{The figure shows the case  $a=1$ corresponding to a waterfall version of natural inflation. We show 1D and 2D 
posterior marginalized distributions of two sampling parameters $\{n_s, r\}$ and derived parameters $\{f, n_{sk}, n_{skk}\}$ using Planck+WMAP dataset (in blue), and the combination of both datasets (PL+WP+BP2, in red).  }}
\label{figHNIa1}
\end{figure}

The case $a=1$ corresponds to a slight generalization of \NI. 
Released from the requirement to end inflation, aided by a waterfall field, we consider the special case of  {\it Waterfall Natural Inflation (WNI)} 
(for a recent study of the consistency of standard {\it NI} with CMB observations of Planck and BICEP2 see \cite{Freese:2014nla}). 
The likelihood contours for this case are presented in Fig.~\ref{figHNIa1}. A certain degeneracy is observed between $r$ and 
the running parameter $n_{sk}$ when analysing the Planck+WP data alone. Such degeneracy is broken when BICEP2 observations are 
added to the dataset. Additionaly, note that the area of overlap of both datasets is marginally smaller than the previous case. 
Note that the preferred value of the symmetry breaking scale $f$ for \WNI~ still lies above unity.

\subsection{Hybrid Natural Inflation with $f=1$} \label{Resultsf} 

\noindent

\begin{figure}[t!]\centering
\includegraphics[width=18cm,height=6.0cm]{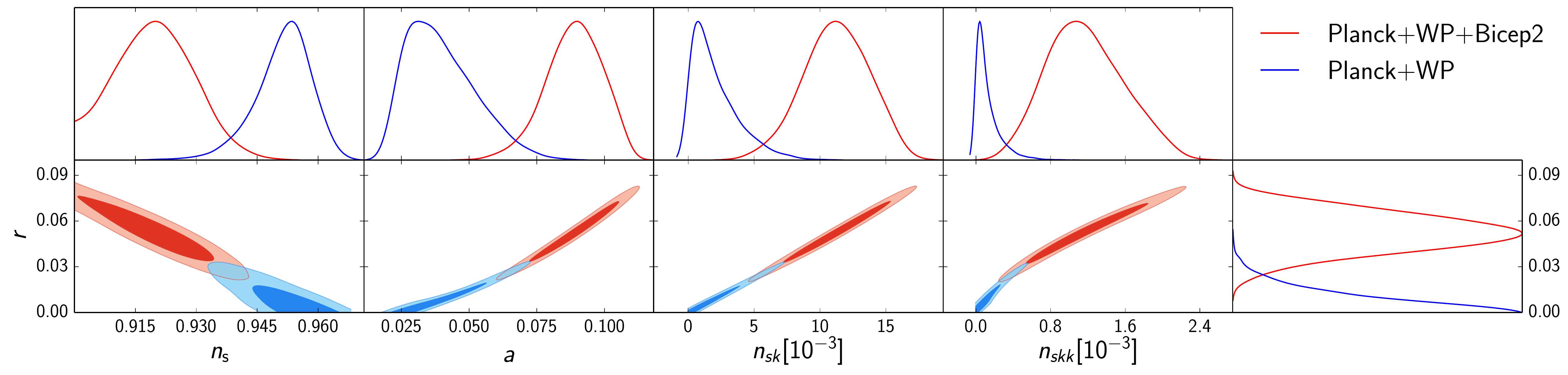}
\caption{\footnotesize{We show 1D and 2D posterior marginalized distributions of two sampling parameters 
$\{n_s, r\}$ and derived parameters $\{a, n_{sk}, n_{skk}\}$ using Planck+WMAP dataset (in blue), and the combination
of both datasets (PL+WP+BP2, in red) for the $f=1$ case. We notice a strong tension between the two data sets.}}
\label{figHNIf1}
\end{figure}

One of the main motivations for introducing the model of {\it HNI} \cite{Ross:2009hg,Ross:2010fg}, where the end of inflation is not triggered by the slow-rolling 
inflaton field but by a second waterfall field, is to release the parameter $f$ from inflationary constraints and to have values equal or less than the Planck scale. Being $f$ the scale of (Nambu-Goldstone) symmetry breaking it is desirable not to surpass the Planck value as occurs in ordinary {\it NI}, where the same field which produces inflation is also in charge of terminating it. 
 What we see from Fig.~\ref{figHNIf1} is that {\it HNI} with $f=1$ is consistent  with Planck+WP data alone but, when taking running through the BICEP2 dataset the overlap is almost vanishing. This evidences a strong tension between the two datasets for the model. Also, a set value for the parameter $f$, means that the single degree of freedom $a$ determines the value of the other parameters. This is evident from the plots which reveal a strong correlation between $r$ and the rest of the spectral parameters. We analyse the consequences of these results for the validity of the models in the next section. 

\begin{figure}[t!]\centering
\includegraphics[width=5.5cm,height=4.5cm]{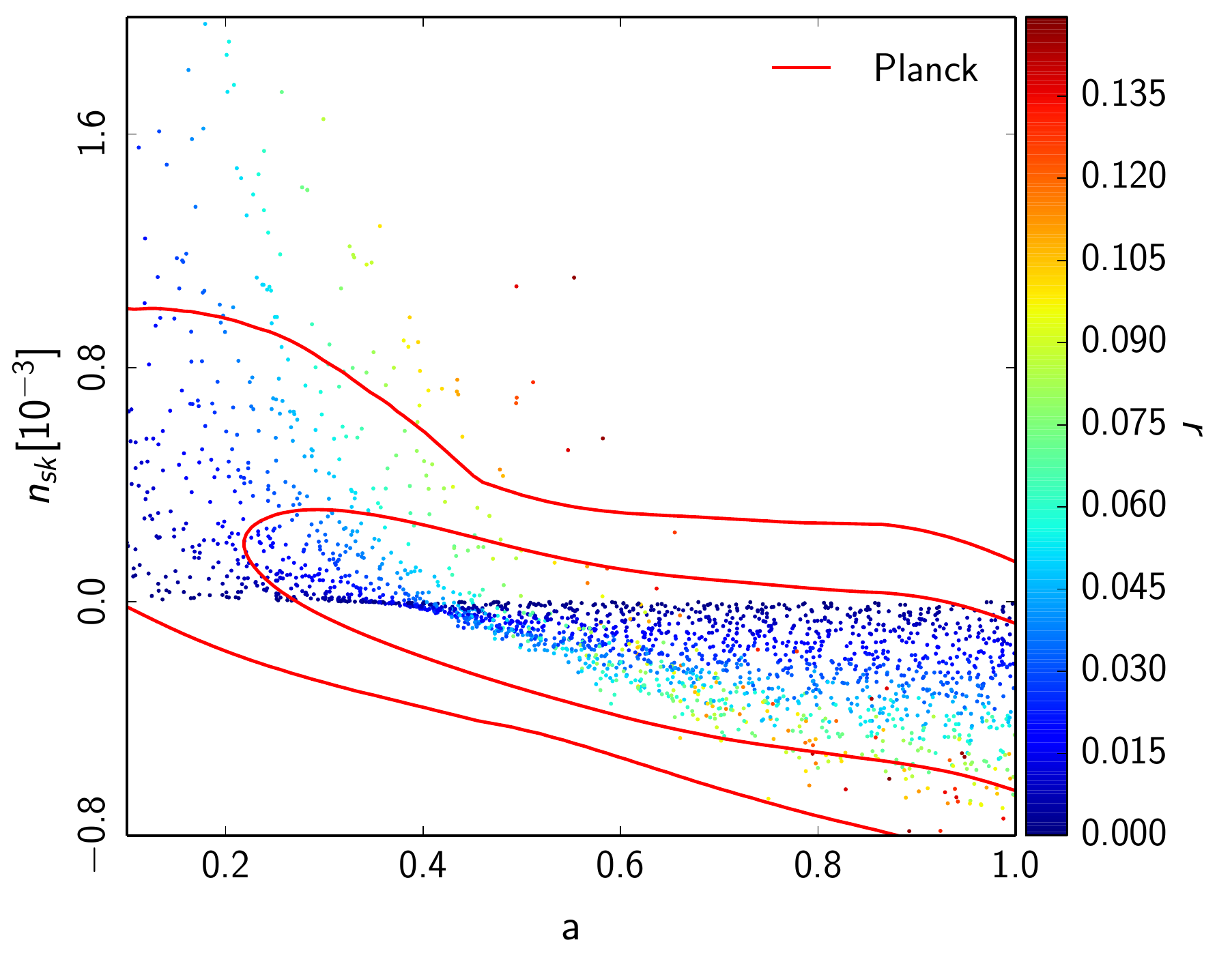}
\includegraphics[width=5.5cm,height=4.5cm]{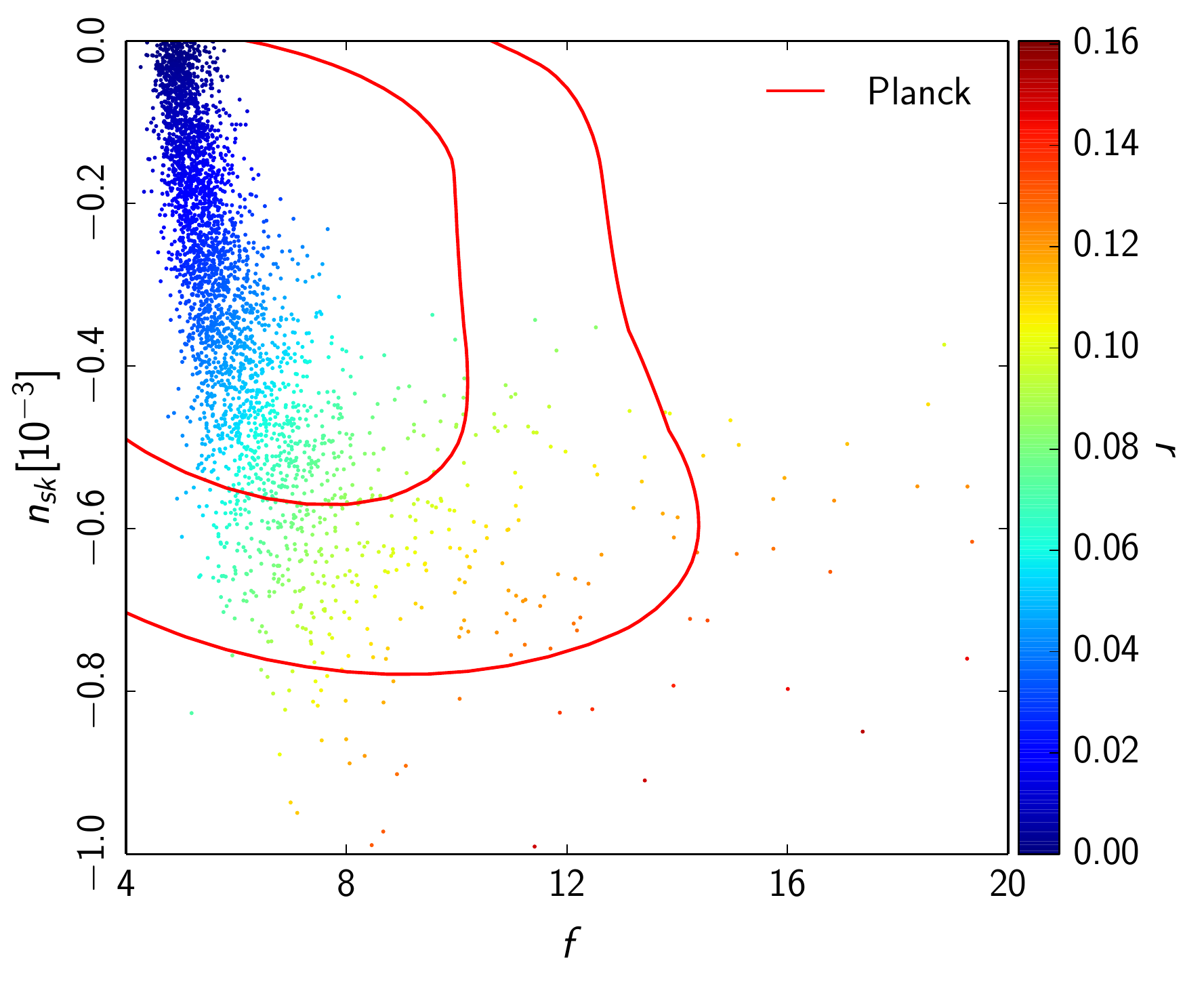}
\includegraphics[width=5.5cm,height=4.5cm]{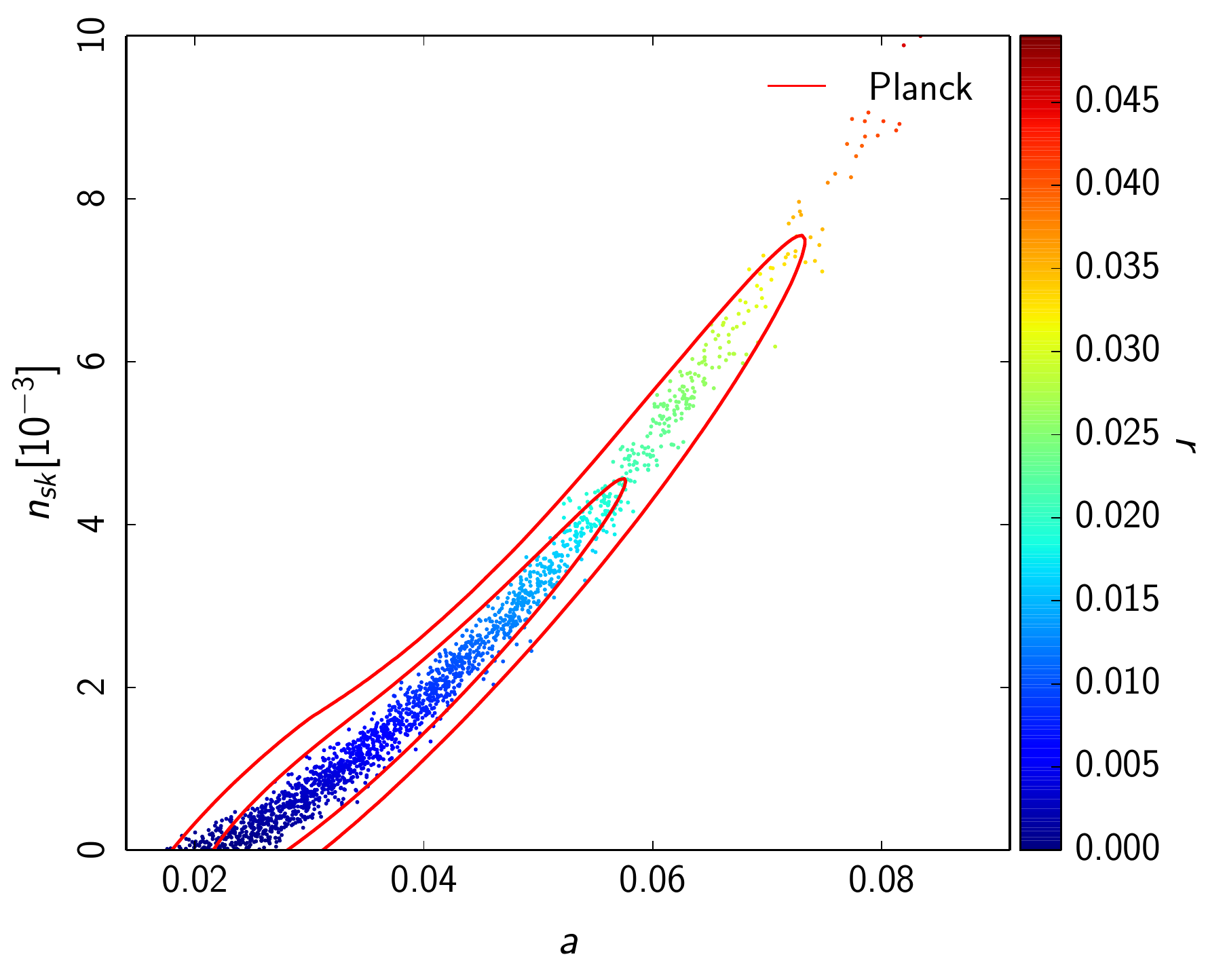}\\
\includegraphics[width=5.5cm,height=4.5cm]{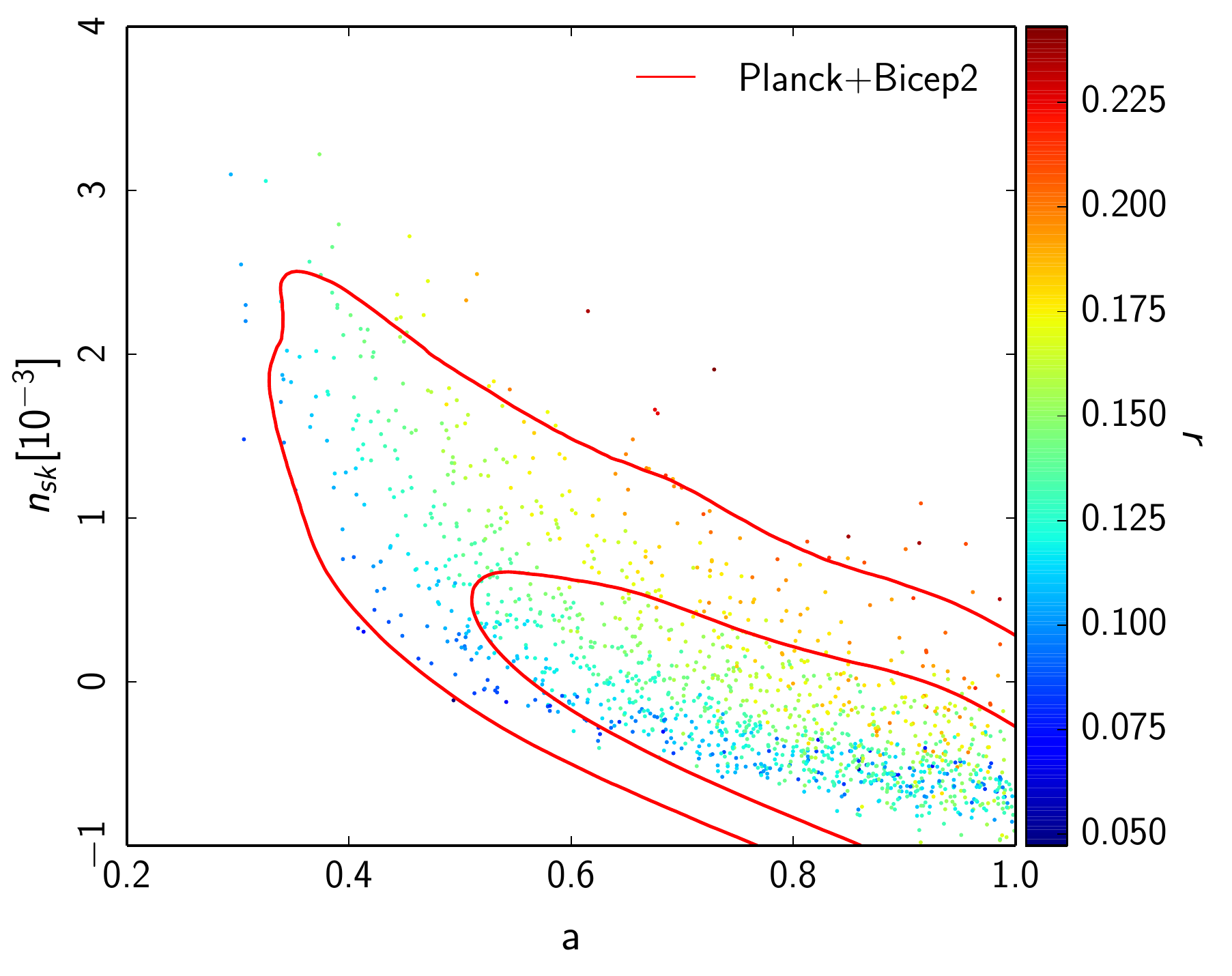}
\includegraphics[width=5.5cm,height=4.5cm]{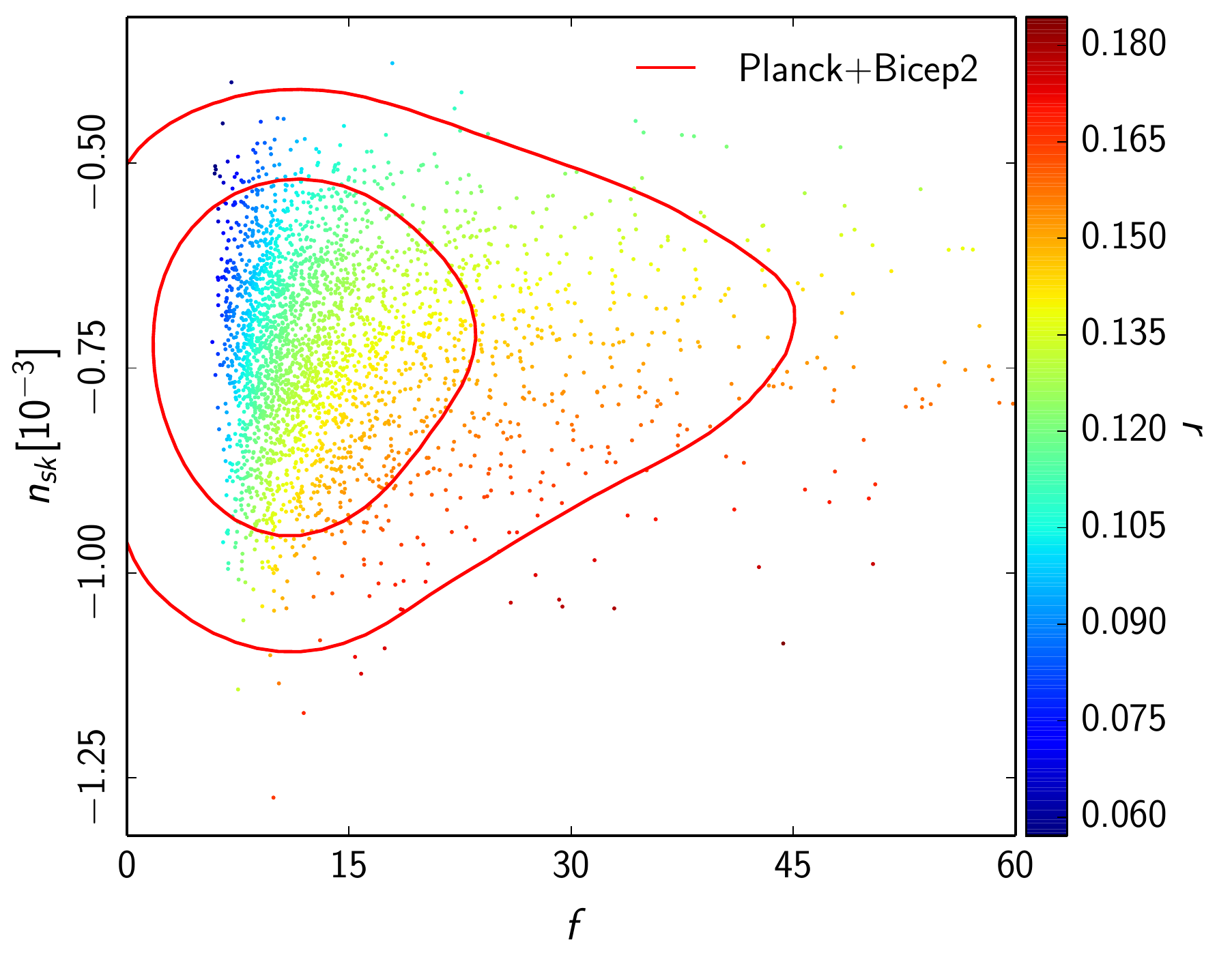}
\includegraphics[width=5.5cm,height=4.5cm]{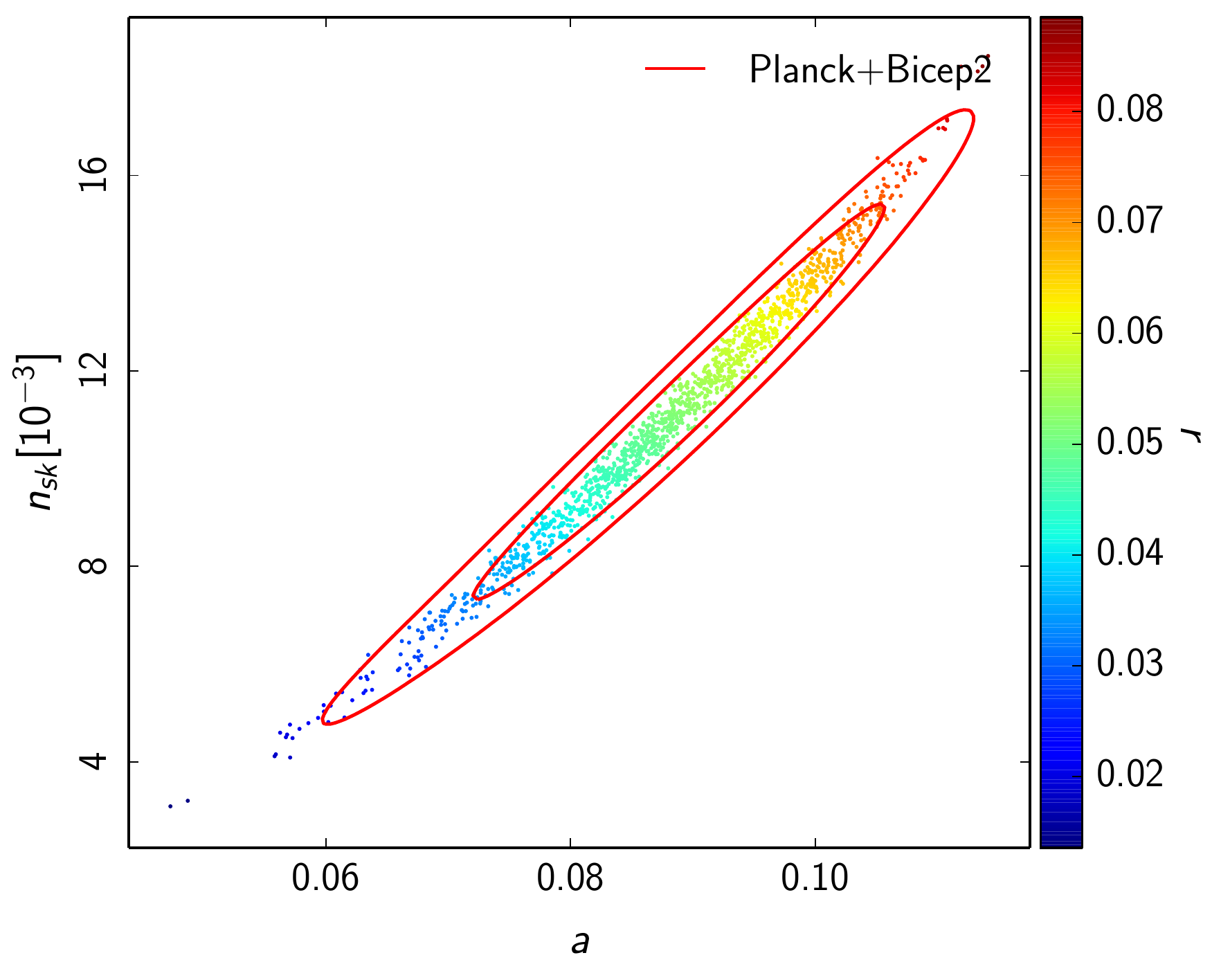}
\caption{\footnotesize{We show a summary of results for the hybrid natural inflation model by means of a 3D posterior distribution. Colour code represents different values of $n_{sk}[10^{-3}]$ 
for Planck+WMAP (above) and for Planck+WMAP+BICEP2 results (below) for the cases (from left to right): general, a=1 and  f=1. The central values of these parameters are presented in Table \ref{table1}}.}
\label{PS1} 
\end{figure}
\begin{table}[htbp]
  \centering
    \begin{tabular}{|c|c|c|c|c|c|}
    \hline
      {\it HNI} & $f$ & $a$ & $r$   & $n_s$ & $n_{sk}$   \\
      \hline
    \textit{Case: general} & $$ & $$ & $$ & $$ & $$     \\
    \hline
    \textit{PL+WP} & $4.5$ & $0.60$ & $<0.09$ & $0.96$ & $-4\times 10^{-5}$     \\
    \hline
    \textit{PL+WP+BP2} & $8.0$ & $>0.42$ & $0.14$ & $0.96$ & $<1.6\times 10^{-3}$     \\
    \hline
  Case: $a=1$ & $$ & $$ & $$   & $$ & $$   \\
      \hline
    \textit{PL+WP} & $6.1$ & $1$ & $<0.1$ & $0.96$ & $-3\times 10^{-4}$      \\
    \hline
    \textit{PL+WP+BP2} & $17.2$ & $1$ & $0.13$ & $0.96$ & $-7\times 10^{-4}$   \\
    \hline
    \textit{Case: $f=1$} & $$ & $$ & $$ & $$ & $$     \\
    \hline
    \textit{PL+WP} & $1$ & $0.04$ & $<0.02$ & $0.95$ & $2\times 10^{-3}$     \\
    \hline
    \textit{PL+WP+BP2} & $1$ & $0.09$ & $0.05$ & $0.92$ & $1\times 10^{-2}$     \\
    \hline
    \end{tabular}%
\bigskip
\caption{We show approximate mean values and bounds for the model parameters $f$ and $a$ (when not fixed) as well as for the tensor $r$, spectral index $n_s$ and running $n_{sk}$ for three cases of hybrid natural inflation using Planck+WMAP data only (PL+WP) and Planck+WMAP+BICEP2 data (PL+WP+BP2). We confirm that for the case of two free parameters $\{a,f\}$, {\it HNI}\, is no real improvement over a {\it Waterfall Natural Inflation} model (i.e. the $a=1$ case). The different values for $r$ in these two cases essentially correspond respectively 
to adjustments in $a$ and $f$, and $f$ alone, which however do not affect the spectral index value. When we fix the scale of symmetry breaking $f$ to one (in Planck units), we see that there is a clear distinction between results. {\it HNI} is consistent with PL+WP data alone, but then becomes disfavoured when PL+WP+BP2 data is considered.}
\label{table1}%
\end{table}%

\section {Consistency checks: $f\leq1$ and bounds on Primordial Black Holes abundance} \label{Analyses}

\noindent

Let us now analyse the results presented above in terms of the self-consistency of the models and the consistency with the constraints imposed by the null observations of primordial black holes. 
From Figs.~\ref{figHNIaf} and \ref{figHNIa1}, we immediately read that the model requires a symmetry breaking parameter $f$ with a value well above the Planck scale which we consider undesirable. Requiring that the scale $f$ is at most $f=1$, in Planck units, we get Fig.~\ref{figHNIf1} where we 
see that \HNI~ is still compatible with the Planck+WP data. When combined with the BICEP2 data however, we find that there is a 
strong tension between the two datasets with an almost vanishing overlap. This is most evident from the combined data plotted in Fig.~\ref{PS1}. To get a handle of these results we present in Table~\ref{table1}
the mean values for the model parameters $f$ and $a$ as well as for the observable tensor $r$, spectral index $n_s$ and running $n_{sk}$ for the three cases of \HNI~ studied here.

When, for the case $f=1$, we combine both datasets there is an additional constraint coming from the possible over-production of primordial black holes (PBHs). The large positive value of the running $n_{sk}$ in Fig.~\ref{figHNIf1} (see also Fig.~\ref{PS}), could invert the red spectrum of scalar perturbations and induce, at small scales, inhomogeneities with amplitudes ruled out by observations. Following the argument of \cite{Kohri:2007qm}, the undesired amplitude for perturbations at which PBHs could be 
over-produced is $\mathcal{P}_s(N= 0)\simeq 10^3$ (see also Refs.~\cite{Josan:2009qn,Carr:2009jm}). Since this is most likely 
to happen toward the end of inflation, when the number of remaining e-folds of inflation is $N \approx 0$, the inflaton field 
has the order of 50 or 60 e-folds to evolve from the initial $\mathcal{P}_s = 10^{-9}$ to the PBHs overproduction value. 
The Taylor expansion of the power spectrum around its observed value at $N_i = 50 \text{ to } 60$ shows that 
\footnote{Note that, to lowest order in slow-roll $d/dN = - d / dk$},
\begin{equation}
\label{ps:expansion}
\ln \left[\frac{\mathcal{P}_s(0)}{\mathcal{P}_s(N_i)}\right] = (n_s- 1) N_i + \frac{1}{2} n_{sk} N_i^2.
\end{equation}

\noindent In writing this equation we have ignored derivatives of the power spectrum at orders higher than two. Such truncated expansions may seem incomplete to evaluate the amplitude of the spectrum, at about 50 e-folds away from observable scales, but they have been successfully employed to constrain undetermined parameters (e.g.~\cite{Carr:1994ar,Kohri:2007qm,Alabidi:2009bk}). 
Our equation \eqref{ps:expansion} is only a first approximation to establish a bound on the undetermined value of the running $n_{sk}$ and therefore to constrain the predictions of \HNI.\footnote{We have checked that the next order term in the expansion of Eq.~\eqref{ps:expansion}, involving the parameter $n_{skk}$, is subdominant with respect to those appearing in the equation.} 

Upon substitution of values of $\mathcal{P}_s(0)$ that over-produce PBHs, we find that the left hand side should not exceed the value of about 14. The right hand side then imposes a constraint on positive values of the running. Substituting the values of  Table~\ref{table1} 
we see that the {\HNI} model with $f = 1$ is restricted to values of the running $n_{sk} < 0.0104$, which 
practically excludes the central value of the parameter adjusting the Planck+WP+BICEP2 data. Note that for this same model, the Planck+WP data yield a smaller value of $n_{sk}\approx 10^{-3}$, which is not excluded by the abundance of PBHs. Our calculation shows that, when the BICEP2 data is interpreted as a signature of primordial gravitational waves, the model of \HNI~ with $f=1$ would be discarded. 

\section{Summary and conclusions} \label{Conclu} 

\noindent

We find that when the scale of symmetry breaking $f$ is left unrestricted, {\it Hybrid Natural Inflation} is not essentialy different from natural inflation both favouring a small running consistent with a vanishing value. The case where $f=1$, however, is consistent with Planck+WP data alone but not when combined with BICEP2 data. This is shown in Fig.~\ref{figHNIf1} evidencing a strong tension between both data sets. Table~\ref{table1} shows that when the Planck+WP+BICEP2 data is used, \HNI~  with $f=1$ is disfavoured since the spectral index starts moving away from 
observed values. At the ultraviolet end of the spectrum, the over-production of PBHs imposes a bound on the value of the running $n_{sk}\lesssim 0.01$. This has also been reported in previous works of single-field  inflation: for running mass inflation \cite{Kohri:2007qm} and for the more general hilltop inflationary models \cite{Alabidi:2009bk}. The common feature shared by these models and \HNI~is that the slope of the inflationary potential flattens toward the end of inflation, allowing for both a large number of e-folds and an enhancement in the amplitude of field fluctuations.

Two comments are in place regarding the validity of constraints derived from BICEP2 and the bounds to PBHs abundance. Recent observations \cite{Adam:2014bub} have raised doubts regarding the cleanliness of the BICEP2 data. Additionally, our results seem to support the argument that BICEP2 data at face value may be incompatible with slow-roll inflation in general \cite{Cortes:2014nqa}. Thus, a conclusive analysis on the feasibility of \HNI~awaits more data.

 On the other hand, recent studies have shown that the formation of PBHs is not solely determined by the amplitude of perturbations. Indeed, not all profiles of overdensities with a common amplitude will get to form PBHs \cite{Polnarev:2006aa,Nakama:2013ica}, and the probability of formation might be significantly reduced, \cite{Hidalgo:2008mv}. On the other hand, the overproduction of PBHs at the end of inflation could trigger the subsequent reheating period \cite{Hidalgo:2011fj}. While this argument requires fine-tuning, it is worth exploring the final stages of inflation since, in addition to the analysis above, a waterfall field, like the one expected to end inflation in the models studied here, is likely to produce PBHs copiously at the end of inflation \cite{Lyth:2012yp}. We leave the study of these possibilities for future work. 

In conclusion, our results show that when we force \HNI~to desirable values of the symmetry breaking parameter, the $f=1$ case is compatible with Planck+WP-only data but not when combined with BICEP2 since a smaller spectral index $n_{s}$ is favoured and over-production of PBHs at the end of inflation occurs. This disagreement is not yet ruling out the $f = 1$ scenario given that the polarisation data from Planck \cite{Adam:2014bub},  indicates that the signal detected by BICEP2 may be completely due to galactic dust. A forthcoming cross-correlation of the data from both experiments will reveal the nature of the BICEP2 observations. 
\section{Acknowledgements}

We gratefully acknowledge support from \textit{Programa de Apoyo a Proyectos de Investigaci\'on e Innovaci\'on 
Tecnol\'ogica} (PAPIIT) UNAM, IN103413-3, \textit{Teor\'ias de Kaluza-Klein, inflaci\'on y perturbaciones gravitacionales} 
and IA101414-1, \textit{Fluctuaciones no-lineales en cosmolog\'{\i}a relativista}.  AHA is grateful to the staff of ICF, 
UNAM and UAM-Iztapalapa for hospitality. MCG acknowledges a scholarship from PI and CONACyT. JAV, GG, AHA and JCH thank SNI for support.

\end{document}